\documentclass[preprint2,longabstract]{aastex}
\usepackage{amsmath}

\slugcomment{Accepted for Publication in the Astrophysical Journal}

\shorttitle{Molecular tracers of accelerated collapse}
\shortauthors{Lintott et al.}

\begin{document}

\title{Molecular abundance ratios as a tracer of accelerated collapse in regions of high mass star 
formation?}

\author{C. J. Lintott, S.Viti, J. M. C. Rawlings and D. A. Williams }
\affil{ Department of Physics and Astronomy, University College London, Gower Street, London, WC1E 6BT, UK}
\email{cjl@star.ucl.ac.uk}

\author{T. W. Hartquist}
\affil{School of Physics and Astronomy, The University of Leeds, Woodhouse Lane, Leeds, LS2 9JT, UK}

\author{P. Caselli}
\affil{ INAF-Osservatorio Astrofisico di Arcetri, Largo E.
Fermi 5, I-50125 Firenze, ITALY}

\author{I. Zinchenko}
\affil{Institute of Applied Physics of the Russian Academy of Sciences, Ulyanova 46, 603950 Nizhny Novgorod, Russia}

\author{P. Myers}
\affil{Harvard-Smithsonian Center for Astrophysics, 60 Garden Street, Cambridge, MA 02138, USA}

\maketitle

\begin{abstract}
Recent observations suggest that the behaviour of tracer species such
as $\mathrm{N_2H^+}$ and $\mathrm{CS}$ is significantly different in regions of high and low mass
star formation. In the latter, $\mathrm{N_2H^+}$ is a good
tracer of mass, while $\mathrm{CS}$ is not. Observations show the reverse to be true in high-mass star formation regions. We use a computational
chemical model to show that the abundances of these and other species may be significantly altered by a period of accelerated collapse in high mass star forming regions. We suggest these results provide a potential explanation of the observations, and make predictions for the behaviour of other species.
\end{abstract}


\section*{Abstract}
Recent observations suggest that the behaviour of tracer species such
as $\mathrm{N_2H^+}$ and $\mathrm{CS}$ is significantly different in regions of high and low mass
star formation. In the latter, $\mathrm{N_2H^+}$ is a good
tracer of mass, while $\mathrm{CS}$ is not. Observations show the reverse to be true in high-mass star formation regions. We use a computational
chemical model to show that the abundances of these and other species may be significantly altered by a period of accelerated collapse in high mass star forming regions. We suggest these results provide a potential explanation of the observations, and make predictions for the behaviour of other species.

\section{Introduction}

Molecular lines are effective probes of physical conditions in star
forming regions. The abundances of molecules and their line profiles
give information on the physical state of the gas and its likely
evolution during the collapse to form a star \citep{b10}. Methods
based on molecular lines have been used together with dust continuum
emission to probe regions of low-mass and of high-mass star formation
\citep{b9}. Recent work has shown that tracer species may be
significantly depleted in the central parts of dense cores. In low-mass star 
forming regions, CS and CO do not trace the gas mass in the
central cores \citep{b1,b3,b11,b12,b2}, but $\mathrm{N_2H^+}$ is an
excellent tracer of the dust continuum emission \citep{b19,b4,b3}. One
possible explanation, therefore, is that the $\mathrm{N_2}$ molecules
which drive the nitrogen chemistry do not freeze-out on to dust grains
as readily as some other species, and the species derived from
$\mathrm{N_2}$, notably $\mathrm{N_2H^+}$, remain present in the gas
to higher densities and trace its mass, as does the dust continuum
emission.

In high-mass star forming regions, however, the situation is
unclear. The epoch most readily studied is the so called hot core
phase, in which a characteristic chemistry signifies the operation of
processes unlike those in quiescent clouds. This chemistry arises from
the evaporation of molecules that had been frozen out on dust grains
during the long cold period of collapse; they are then released back
to the gas phase when the newly formed massive star warms its
environment so that the ices sublimate.

Recent molecular line observations \citep{b8,b9,b14} of high-mass-star forming cores
show differences between the CS and $\mathrm{N_2H^+}$
distributions.  In a survey of sources in regions of high-mass star
formation, Pirogov et al. find that the CS and dust emission
distributions are very similar, and differ significantly in many cases from the
$\mathrm{N_2H^+}$ distributions; the $\mathrm{N_2H^+}$ abundance falls
near massive young stellar objects, on scales up to a parsec where the temperature, relative to the rest of the cloud and to the CS peak,
does not change significantly. The
variation in relative intensity can be about one order of magnitude, larger than could be
accounted for by excitation or opacity effects, and is seen in approximately one-third of the 
sources. {\bf A detailed study} is, however, expected to reveal significant differences in the majority of
the sources. Zinchenko et al. have
also discussed the behaviour of some other molecules in dense cores in
high-mass star forming regions. In summary, in contrast to its
behaviour in low-mass star-forming regions where $\mathrm{N_2H^+}$
traces mass very well, in regions of high-mass star formation
$\mathrm{N_2H^+}$ is a poor tracer of mass.

Here, we focus on the exciting new results concerning CS and
$\mathrm{N_2H^+}$. Why is there a complete reversal of the behaviour
between regions of formation of high-mass stars (in which CS is
apparently a good tracer of core mass) and of low mass stars (in which
$\mathrm{N_2H^+}$ is apparently a good tracer of core mass)? We offer
a possible explanation in terms of the differing dynamical state of
the two types of region, and with a detailed computational model of high-mass star formation we
obtain results that are consistent with the observations.

\section{Observations}

Here, we briefly review the observations reported in detail elsewhere
 \citep{b8,b9,b16,b17,b14}. $\mathrm{N_2H^+}$(1-0) emission was observed in 2000 -
2001 at the Onsala 20 m telescope and at SEST. CS(2-1) has also been observed at SEST and Onsala. CS(5-4) emission was
observed at SEST and also with $\mathrm{C^{34}S}$(5-4) at the NRAO 12
m telescope. 35 sources were observed in the Onsala/SEST $\mathrm{N_2H^+}$(1-0) surveys, and
about 20 of these were later observed in dust continuum emission at SEST
using the SIMBA bolometer array and at IRAM 30m with the MAMBO array. The CS is found to correlate well
with the dust emission, while $\mathrm{N_2H^+}$ and CS appear to be
poorly correlated in many cases. Several sources were mapped also in HCN,
HNC, HCO$^+$, CH$_3$OH, and several other species. The gas kinetic temperatures were
estimated from CH$_3$C$_2$H observations. The most complete data
sets were obtained for S255 and S187. Inferred temperatures for
S255 from the CH$_3$C$_2$H observations are about 35 K, both at
the CS and $\mathrm{N_2H^+}$ emission peaks, somewhat warmer than
cold cloud conditions.

As an example, in figure 1 we show $\mathrm{C^{34}S\left(2-1\right)}$ and $\mathrm{N_2H^+}$(1-0) maps of S255 overlaid on the map of the dust continuum emission at 1.3mm. Both maps were obtained at Onsala with Nyquist sampling (HPBW/2=20'' grid spacing). The  $\mathrm{C^{34}S\left(2-1\right)}$ emission is presumed to be optically thin. The peak optical depth for $\mathrm{N_2H^+}$(1-0) in S255 is $\approx$ 0.5 (Pirogov et al. 2003).

The dust continuum data shows two peaks, almost equal in intensity. The $\mathrm{C^{34}S\left(2-1\right)}$ distribution is similar, although the emission at the southern peak is somewhat stronger. The $\mathrm{N_2H^+}$ is very different; the northern component is much stronger in this line than the southern one. 

The nature of these two components is different. The southern one is associated with a luminous cluster of IR sources, whereas toward the northern one an ultracompact H II region (G192.58-0.04) was detected. This object is extremely red in the mid-IR band and Crowther \& Conti, and Mezger et al. derived almost exactly the same dust masses and temperatures for both components from 1.3 mm and $350\mu$ m observations. Their dust temperatures are very close to the gas kinetic temperature referred to above. We interpret the northern source as a HII region, and the southern as a young high-mass protostellar candidate. It is this latter class of object that we model in this paper.

\section{Rapid collapse of cores in high-mass star forming regions}

While cores in low-mass star forming regions are generally quiescent
\citep{b7,b6}, regions of high-mass star formation are dynamically
active. We therefore explore the possibility that additional external
pressures arising from this dynamical activity acting on a core can
enhance the rate of collapse above the natural free-fall rate. We do
not consider the possible mechanisms for this enhanced rate of
collapse in detail. In fact, such an enhancement could be due to compression by
shocks that are sufficiently weak that they have little effect on the
temperature. Free-fall is conventionally assumed to be the limiting
rate in regions of low-mass star formation. For convenience, we
use the free-fall collapse law described in Rawlings et al. (1992) and
we specify the enhancement of collapse in high-mass star forming
regions by setting a larger value of the inverse time constant in the
free-fall collapse equation which relates density and time. The change in number density, $n$, as
collapse proceeds from initital density $n_0$ to a final density $n_f$ is 
then given by

\begin{equation}
\frac{\mathrm{d}n}{\mathrm{d}t}  = B \left(\frac{n^4}{n_0}\right)^{\frac{1}{3}}\left( 24\pi Gm_{H} n_{0} \left[ \left(\frac{n}{n_0}\right)^{\frac{1}{3}}-1\right]\right)^{\frac{1}{2}}
\end{equation}

where $t$, $G$ and $m_{H}$ are time, the gravitational constant and the mass of the hydrogen atom respectively. Values of $B$ which are greater than 1 allow accelerated collapse to be simulated.

 Though
the collapse law is in reality somewhat different, the adoption of the
enhanced free-fall description allows us to explore the qualitative
response of the chemistry to more rapid collapse. The timescale of the
collapse compared to the timescale of freeze-out is more important for
the chemical results than the other details of the collapse
history. \emph{The effect of an enhanced collapse rate should be that high
gas densities would be achieved rapidly, before the effects of
freeze-out dominate the chemistry.}

The primary formation route for $\mathrm{N_2H^+}$ is through the reaction between $\mathrm{H_3^+}$ and $\mathrm{N_2}$ which also produces molecular hydrogen. {\bf The $\mathrm{N_2H^+}$ molecule} reacts with species such as $\mathrm{H_2O}$, C, O or CO to form $\mathrm{N_2}$ and a positive ion. It is easy to see, therefore, that the abundance in the gas phase of these removal agents will be critical in determining the abundance of $\mathrm{N_2H^+}$ present.

In particular, we explore the
hypothesis that high densities would be reached before molecules that
are important as $\mathrm{N_2H^+}$ removal agents are
fully depleted in the gas phase. If this is indeed the case,
then $\mathrm{N_2H^+}$ abundances should be reduced in such
circumstances, while the higher gas density promotes gas phase
chemistry producing, in particular, CS, through reactions such as the interaction
of $\mathrm{He^+}$ and HCS, or between atomic carbon and diatomic molecules such
as SO. In the canonical model with free fall collapse, molecules such as CS freeze out of the gas
phase before being destroyed in further reactions.

We expect, therefore, that in accelerated collapse, CS should be enhanced while
$\mathrm{N_2H^+}$ is reduced, compared to canonical calculations, and
use this prediction to test our hypothesis. We use a chemical
model similar to that used in Viti \& Williams (1999) to describe the
formation and chemistry of high mass star formation. The model in Viti
\& Williams (1999) is a two stage calculation: the first stage starts
from a fairly diffuse ($\sim$ 300 cm$^{-3}$) medium in atomic form
(apart from a fraction of hydrogen in H$_2$), and undergoes a
 collapse until densities typical of hot cores are reached
($\sim$ 10$^7$ cm$^{-3}$). During this phase the temperature is 10K. The collapse follows the single-point
free-fall evolution according to equation 1 (as used by Rawlings et al. 1992). During this
time, atoms and molecules from the gas freeze onto the grains and they
hydrogenate where possible. The advantage of this approach is that the
ice composition is \emph{not} assumed but it is self-consistently derived by a time dependent
computation of the chemical evolution of the gas/dust interaction
process. 

The adopted initial abundances are 1.0, 0.075, 4.45$\times$10$^{-4}$,
1.79$\times$10$^{-4}$, 8.52$\times$10$^{-5}$, 1.43$\times$10$^{-6}$ and
5.12$\times$10$^{-6}$ respectively, for H, He, O, C, N, S and Mg. The
chemical network is taken from the UMIST rate 99 database \footnote{www.rate99.co.uk}
(Le Teuff, Millar \& Markwick 2000). We follow the
chemical evolution of 168 species involved in 1857 gas-phase and grain
reactions.  {\bf At the end of Stage I $\sim$ 99 per cent of the gas is frozen out.}

The second stage follows the chemical evolution of the
remnant core, after the protostar is born. However, for most of our
runs, we only make use of Stage I, as we are interested in the
pre-stellar evolution of the core, given that all analyzed objects show temperatures around 30 K.  We have run several models
exploring the effect of a two-phase collapse where we employ a
free-fall collapse up to a critical density, $n_{crit}$ (a free
parameter), followed by collapse at an enhanced rate (up to 4 times
the free-fall rate, {\bf i.e.   $B>1$ in} equation 1) until a final density of 10$^7$ cm$^{-3}$ is
reached.

For some of our models, we also run Stage II, as described in a
recent development of the Viti \& Williams (1999) model (Viti et
al. 2004): we simulate the effect of the presence of an infrared
source in the centre of the core or in its vicinity by subjecting the
core to a time-dependent increase in the gas and dust temperature up
to 50 K, followed by partial evaporation of grains, according to the
prescription contained in Viti et al. (2004). In these runs, the temperature increase begins once a density $n_T$ is reached.

The parameter choices for our calculations are given in Table 1.

We find, as expected, that enhancing the collapse rate above free-fall
increases the CS abundance and reduces the $\mathrm{N_2H^+}$
abundance. The shift in each case is about half of one order of
magnitude. The results for collapse rate enhancements differ
significantly from the results for free-fall, but not dramatically for
enhancements by factors of 2 - 4 or for accelerations triggered at
different densities. In Fig. 2 we show the ratio of the CS and
$\mathrm{N_2H^+}$ abundances as a function of time for the cases $B =
1$ (free-fall), and $B = 2 - 4$ (enhanced collapse rate).  There is a
substantial enhancement in the ratio in cases in which the collapse
rate is enhanced. The enhancement is not observed when the freeze-out
of molecules is removed from the simulation, supporting the hypothesis that the
change in abundance observed is associated with freeze-out. In
order to ensure that our results could not be replicated simply by
allowing complete depletion, a run was completed in which the rate of collapse
was decelerated by a factor of 10. In this case the $\mathrm{N_2H^+}$
abundance was found to drop rapidly to zero before depletion is
significantly advanced; therefore the observations cannot be explained
by decelerated collapse.

We have investigated the effect of heating to the observed
temperatures of 35-50K following the method of Viti et al., where the
temperature of the core is increased monotonically following a power
law and where the evaporation of a fraction of mantle species $X$ (in
a single step) occurs when the temperature for a particular desorption
event is reached
\citep{b5}. Heating was initiated in the final stages of
collapse; the results were insensitive to the density at which
temperature increase began. Figure 3 shows that the increase in CS and
decrease in $\mathrm{N_2H^+}$ abundances are not significantly
affected by the incorporation of heating into the simulation.

\section{Discussion and Conclusions}

Recent observational data on cores in regions of high-mass star
formation show the exciting and unexpected results opposite in
character to that found in regions of low-mass star formation. We have
suggested that the enhancement of CS and reduction in
$\mathrm{N_2H^+}$ abundance found in regions of high-mass star
formation may be related to the high dynamical activity in these
regions which could enhance the rate of collapse of cores above the
free-fall rate. Consequently, high gas densities would be achieved
before freeze-out had removed the molecules responsible for
$\mathrm{N_2H^+}$ loss, while the high densities promote CS
formation. We have demonstrated using a computational chemical model
that the predicted effects are indeed recovered for plausible
parameter choices. We have shown that these results are not
particularly sensitive to the particular enhancement rate adopted,
above a factor of twice the free-fall rate. The observational results
also give some information about variations of other molecules with
respect to high density tracers. Our model confirms that HCN, for
example, is similar to CS in its distribution, as found by Pirogov et
al. However, our model predicts SO would have a distribution similar
to $\mathrm{N_2H^+}$, whereas the observations of S255 show it to be
much closer to CS. (See Table 2). {\bf As yet, data on the distribution of SO abundance is available 
only for S255 and so it is not clear whether this result is significant or restricted to this
source.} As with most molecules, the effect of accelerated collapse on SO will be both to increase
the rate of production and to prevent the removal
of other possible reaction partners from the gas phase. This discrepancy between theory and
observation for SO suggests that our classification of molecules into two classes of species,
each of which is primarily affected by only one of these mechanisms, is over-simplistic
for molecules at the centre of complex reaction networks. Should further SO observations confirm
the S255 results, a detailed analysis of the behaviour of sulfur-bearing species in our
model will be necessary.

 The temperatures found in these cores are somewhat warmer than 
 normally found in dense cores. We have examined the results from
 computational chemical models in which the initial warming stages of
 the hot core phase is permitted, up to temperatures of around 30 - 50
 K, according to the prescription contained in Viti et al.(2004) The results
 reported here are not significantly changed.

We conclude that the results of the model investigated are consistent
with the rather limited observational data.  This first investigation
of the effects of accelerated collapse on the chemical abundances of
species in high mass star formation regions suggests a possible
explanation of the observations considered, and provides an important
route for future observational and theoretical work.
 
\section*{Acknowledgments}
C.J.L. is supported by a PPARC studentship. SV acknowledges individual financial support from a PPARC
Advanced Fellowship. D.A.W. thanks the Leverhulme Trust for the Leverhulme Emeritus Award.
I.Z. was partly supported by the Russian Foundation for Basic
Research grant 03-02-16307.  P.C. acknowledges support
from the MIUR project "Dust and Molecules in Astrophysical Environments"

\clearpage

\begin{figure}
\includegraphics[angle=270,width=0.20\textwidth]{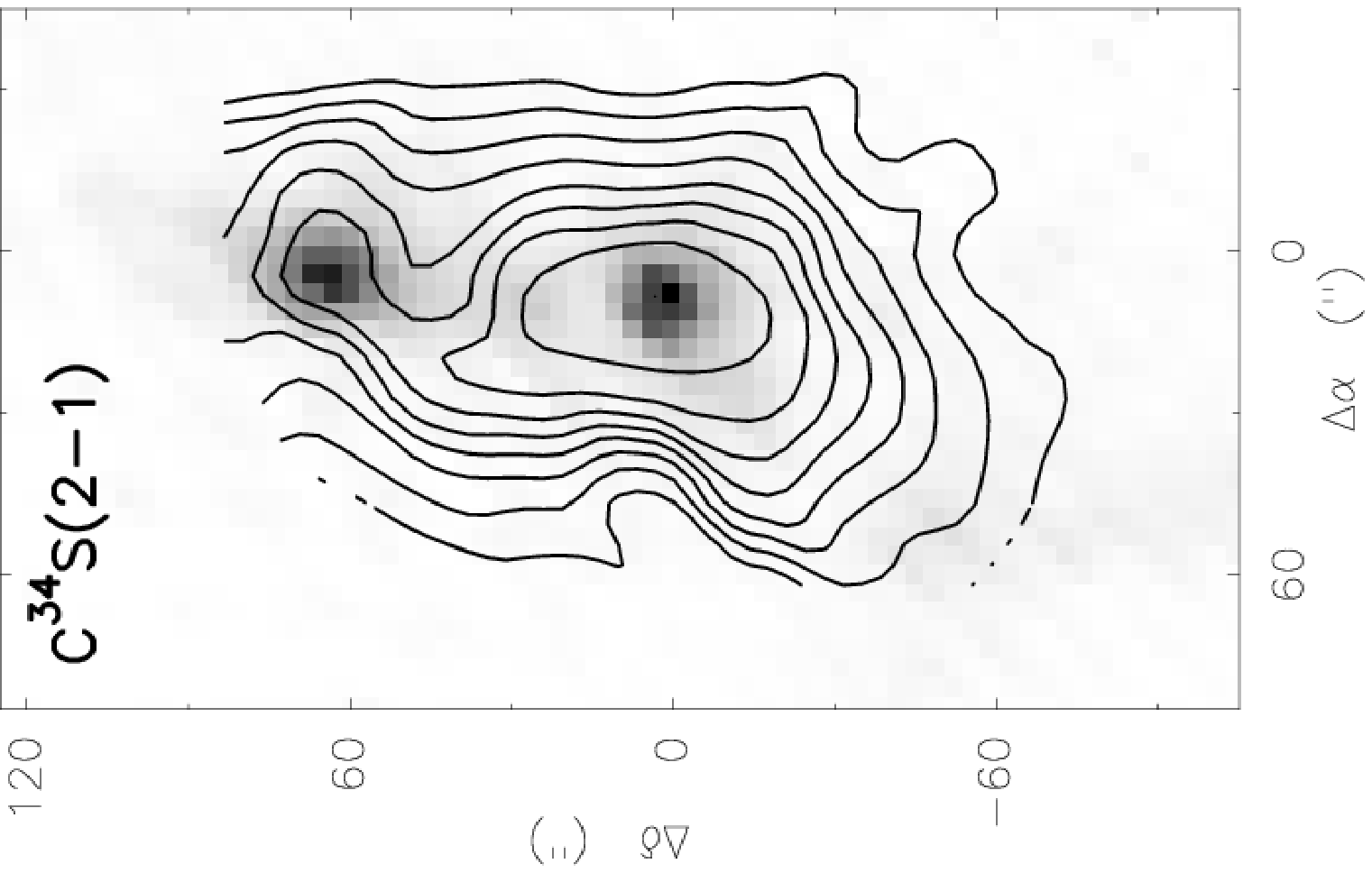}
\includegraphics[angle=270,width=0.20\textwidth]{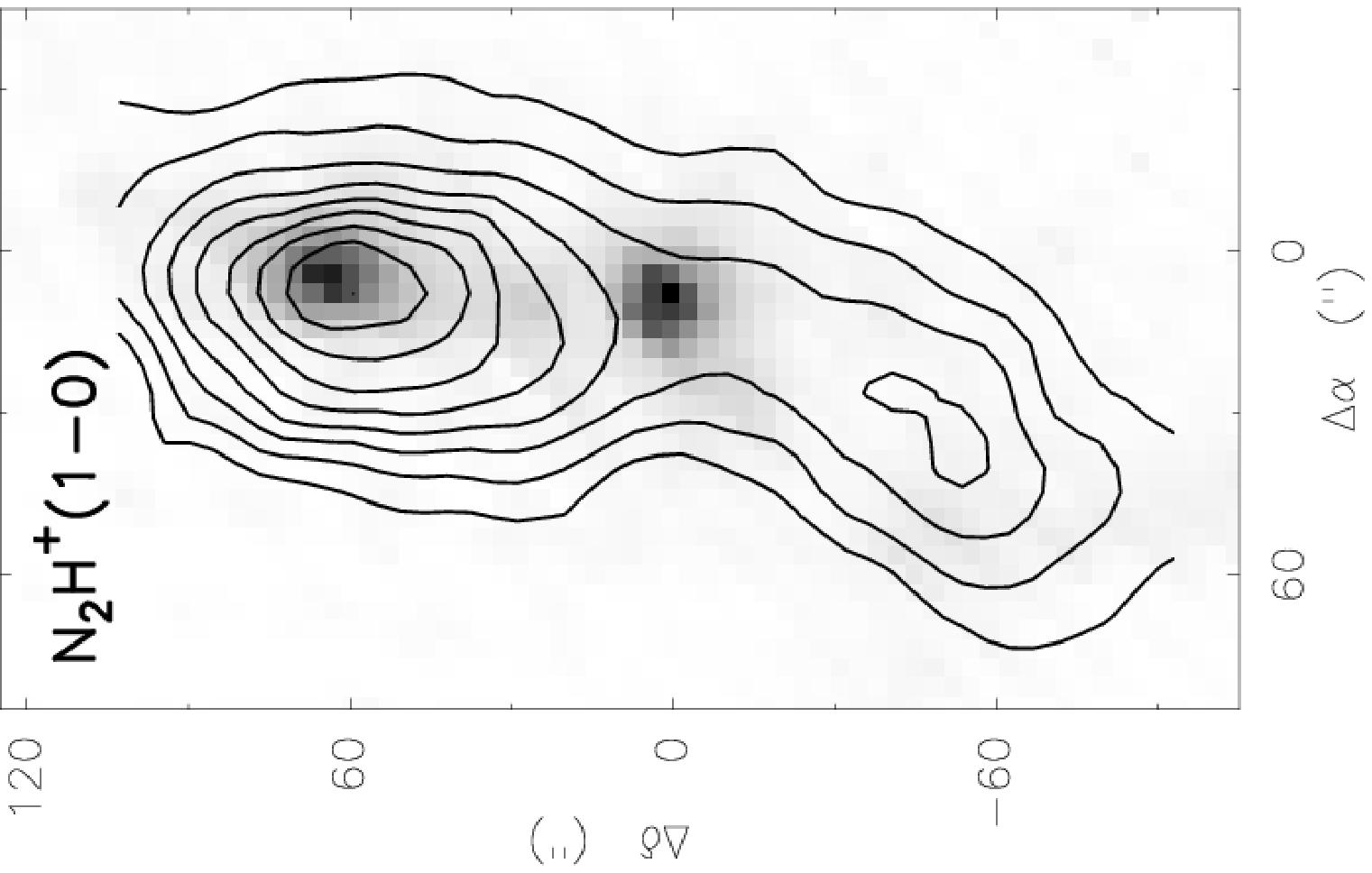}
\caption{Observed contours of $\mathrm{C^{34}S}$(2-1) and $\mathrm{N_2H^{+}}$(1-0) in S255, overlaid on 1.2mm continuum map obtained from IRAM. $\mathrm{C^{34}S}$, which we presume is optically thin, shows a clear correlation with the dust which is absent in $\mathrm{N_{2}H^{+}}$. The model presented in this paper attempts to account for this observed difference in distribution.} 
\end{figure}

\begin{figure}
\includegraphics[width=0.5\textwidth]{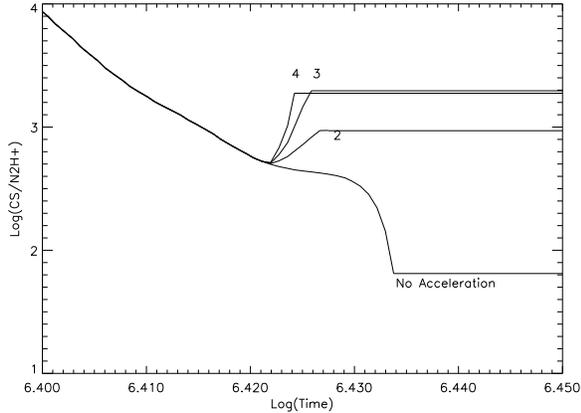}
\caption{Evolution of the ratio of CS to $\mathrm{N_2H^+}$ abundances during
  collapse. Once a critical number density of $10^5 \mathrm{cm^{-3}}$ was
  achieved, three of the four runs (A,B,C \& D in Table 1) shown incorporated a collapse
  accelerated by the factor shown (2, 3 or 4 times the free-fall
  rate). The acceleration increased the abundance of CS and
  decreased the $\mathrm{N_2H^+}$ abundance by approximately an order of
  magnitude; we conclude that this provides a possible explanation of the
  recent observations of high-mass cores.}
\end{figure}

\begin{figure}
\includegraphics[width=0.5\textwidth]{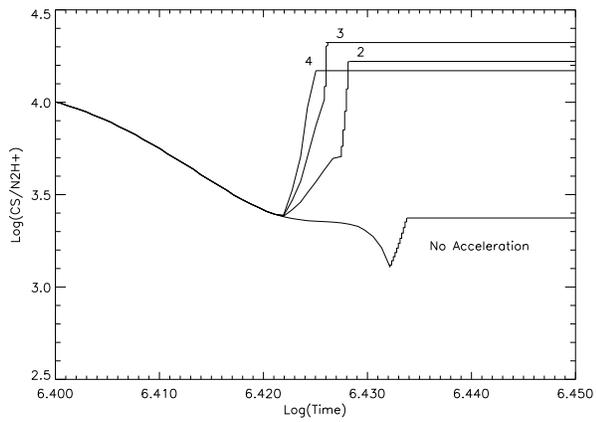}
\caption{Evolution of CS to $\mathrm{N_2H^+}$ abundances in runs F,G,H \& I incorporating an increase 
of temperature to a maximum of 50 K, in the later stages of the collapse.}
\end{figure}

\clearpage

\begin{table}
\begin{tabular}{|c|c|c|c|c|}
\hline
Run & $n_{crit}$ & $B_{new}$ & Temp. Rise? & $n_{T}$\\
\hline
A & $10^5$ & 1.0 & No & N/A\\
\hline
B & $10^5$ & 2.0 & No & N/A \\
\hline
C & $10^5$ & 3.0 & No & N/A \\
\hline
D & $10^5$ & 4.0 & No & N/A \\
\hline
E & $10^5$ & 2.0 & Yes & $10^6$ \\
\hline
F & $10^5$ & 1.0 & Yes & $10^4$ \\
\hline
G & $10^5$ & 2.0 & Yes & $10^4$ \\
\hline
H & $10^5$ & 3.0 & Yes & $10^4$ \\
\hline
I & $10^5$ & 4.0 & Yes & $10^4$ \\
\hline
J & $10^3$ & 2.0 & No & N/A \\
\hline
K & $10^4$ & 2.0 & No & N/A \\
\hline
L & $10^6$ & 2.0 & No & N/A \\
\hline
M &$10^5$ & 0.1 & Yes &$10^6$\\
\hline
\end{tabular}
\caption{For each run, collapse occurs under free-fall until number density $n_{crit}$ is reached, when an acceleration by a factor $B_{new}$ is applied. For some runs, a temperature increase to 50K was initiated at a density $n_T$ as described in the main text.}
\end{table}

\clearpage

\begin{table}
\begin{tabular}{|c|c|}
\hline
$\mathrm{N_2H^+}$-like & CS-like \\
\hline
$\mathrm{NH_3}$ & HCN \\
$\mathrm{HCO^+}$ & OCS \\
$\mathrm{H_2S}$ & \\
SO & \\
$\mathrm{SO_2}$ & \\
\hline
\end{tabular}
\caption{Examples of model predictions of species with abundances which behave like either $\mathrm{N_2H^+}$ or CS.}
\end{table}

\end{document}